\documentclass[journal]{IEEEtran}
\ifCLASSINFOpdf

\else

\fi

\usepackage{CJK}
\usepackage{algorithm}
\usepackage{algorithmic}
\usepackage{amsmath}
\usepackage{amssymb}
\usepackage{graphicx}
\usepackage{epstopdf}
\usepackage{multirow}
\usepackage{stfloats}
\usepackage{cite}
\usepackage{color}
\usepackage[normalem]{ulem}
\usepackage{arydshln}
\usepackage{float}
\usepackage{enumerate}
\usepackage{bbm,pifont}
\usepackage{pstool}
\ifCLASSOPTIONcompsoc
\usepackage[caption=false, font=normalsize, labelfont=sf, textfont=sf]{subfig}
\else
\usepackage[caption=false, font=footnotesize]{subfig}
\fi
\begin{document}

\title{Toward Smart Security Enhancement of Federated Learning Networks}

\author{Junjie Tan, Ying-Chang Liang, Nguyen Cong Luong, and Dusit Niyato
\thanks{The paper has been accepted by IEEE Network, pending for publication. Junjie Tan and Ying-Chang Liang are with University of Electronic Science and Technology of China; Nguyen Cong Luong is with PHENIKAA University; Dusit Niyato is with Nanyang Technological University. (Corresponding Author: Ying-Chang Liang)}
}
\maketitle

\begin{abstract}
As traditional centralized learning networks (CLNs) are facing increasing challenges in terms of privacy preservation, communication overheads, and scalability, federated learning networks (FLNs) have been proposed as a promising alternative paradigm to support the training of machine learning (ML) models.
In contrast to the centralized data storage and processing in CLNs, FLNs exploit a number of edge devices (EDs) to store data and perform training distributively.
In this way, the EDs in FLNs can keep training data locally, which preserves privacy and reduces communication overheads.
However, since the model training within FLNs relies on the contribution of all EDs, the training process can be disrupted if some of the  EDs upload incorrect or falsified training results, i.e., poisoning attacks.
In this paper, we review the vulnerabilities of FLNs, and particularly give an overview of poisoning attacks and mainstream countermeasures.
Nevertheless, the existing countermeasures can only provide passive protection and fail to consider the training fees paid for the contributions of the EDs, resulting in a unnecessarily high training cost.
Hence, we present a smart security enhancement framework for FLNs.
In particular, a verify-before-aggregate (VBA) procedure is developed to identify and remove the non-benign training results from the EDs. Afterward, deep reinforcement learning (DRL) is applied to learn the behaving patterns of the EDs and to actively select the EDs that can provide benign training results and charge low training fees.
Simulation results reveal that the proposed framework can protect FLNs effectively and efficiently.
\end{abstract}


\section{Introduction}\label{sec:introduction}

Nowadays, tens of billions of connected devices in the world are generating an unprecedentedly huge amount of data. Due to the data-driven nature, \emph{machine learning} (ML) has benefited greatly from the data explosion and becomes a vital enabler in many fields, such as computer vision, autonomous cars, and communications \cite{luong2019applications}. The core of ML is about training, i.e., to establish and optimize an ML model, e.g., \emph{deep neural network} (DNN), to seek the relationship contained in the training data, after which the trained ML model can make prediction or decision-making accurately. In traditional ML paradigms, ML models are trained within a \emph{centralized learning network} (CLN), where a server collects and stores all training data into a centralized dataset. However, the collection of raw data not only scarifies privacy but also places heavy burdens on communication infrastructures. Moreover, the centralized data storage and processing are also hardly scalable to the ever-increasing data.

To overcome the challenges, \emph{federated learning} (FL) has been proposed as a favorable alternative of traditional ML paradigms, transforming CLNs into \emph{federated learning networks} (FLNs) \cite{yang2019federated}. In an FLN, multiple \emph{edge devices} (EDs), e.g., smartphones, train an ML model collaboratively in a distributed manner under the coordination of a server while keeping training data locally \cite{mcmahan2016communication}. Specifically, the EDs use their own dataset to perform local training parallelly and upload the results, called model updates, to the server, and the server aggregates the received results to update the ML model. After the repeated interactions between the EDs and the server, the ML model can achieve a satisfactory accuracy, indicating the completion of training.

Compared with CLNs, FLNs have many advantages. On the one hand, the EDs do not need to upload raw data, which avoids privacy concerns and reduces communication overheads. On the other hand, FLNs exploit the EDs to store and process data parallelly, which can benefit from the trends toward massive connected devices and the continuously enhanced storage and computation capabilities in each individual device. Nevertheless, as the training process utilizes the model updates from the EDs, FLNs are susceptible to poisoning attacks. In particular, the malicious EDs can falsify and upload poisoned model updates to the server. Besides, attackers can hijack the model updates transmitted over insecure connections. As a result, the server receives and uses poisoned model updates to obtain a tampered ML model.

In this paper, we first give a brief overview of FLNs and highlight their vulnerabilities to poisoning attacks. After that, we present a summary of the potential poisoning attacks on FLNs and mainstream countermeasures. By analyzing the existing countermeasures, we find that those methods can only provide FLNs with passive protection by means of removing or devaluing part of the model updates received at the server. Consequently, the existing countermeasures have a low utilization of model updates. The problem becomes even more severe in the FLNs with incentive mechanisms, i.e., the EDs charge certain training fees for contributing model updates, incurring an unnecessarily high training cost. Therefore, we propose a smart security enhancement framework to address the issue. In particular, we develop a \emph{verify-before-aggregate} (VBA) procedure to enable the server to identify and remove poisoned model updates. Then, \emph{deep reinforcement learning} (DRL) \cite{mnih2015human} is applied to learn the behaving patterns of the EDs, which are typically determined by attackers and cannot be known to the server, from historical identification results. With the learnt knowledge, DRL allows the server to actively select the EDs that can provide benign model updates at low training fees. Simulation results demonstrate that the proposed framework can offer effective and efficient protection to FLNs.

\section{Fundamentals of Federated Learning Networks}\label{sec:FL_fundamental}

\subsection{Components and Functions of FLNs}\label{sec:FL_basic}

\begin{figure*} [t]
\centering
  \includegraphics[width=16cm]{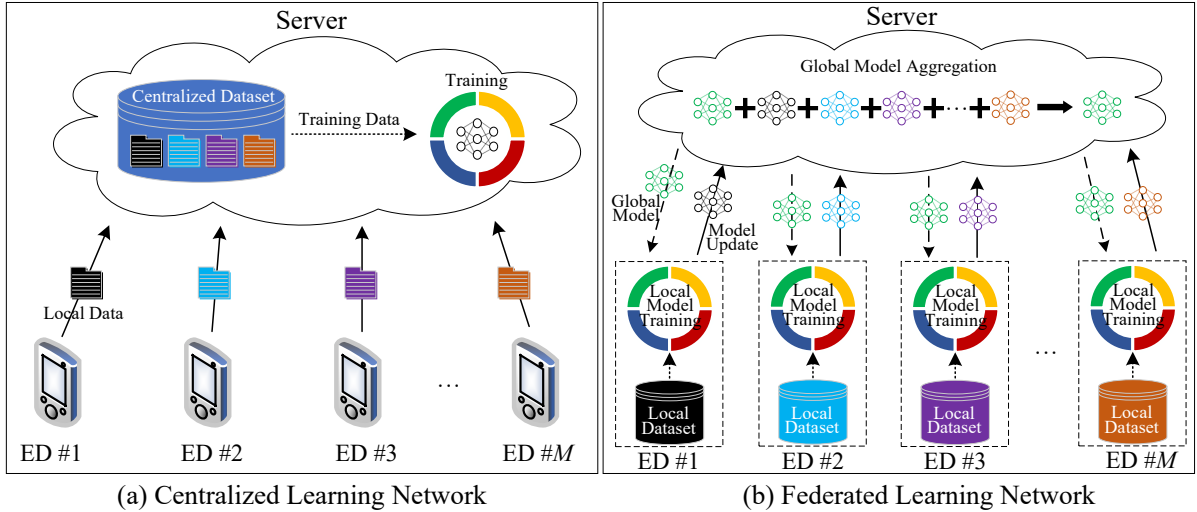}
  \caption{Illustrations of (a) system model of a typical CLN, (b) system model of a typical FLN.}\label{fig:clnfln}
\end{figure*}

Conventionally, ML models are trained centrally, assuming that training data are fully available and stored in a centralized dataset.
Thus, CLNs are designed to support the centralized training. Fig. \ref{fig:clnfln}a shows a typical CLN, which consists of a server and several EDs. Particularly, the EDs upload their local data to the server to build the centralized training dataset, with which the server trains ML models by using \emph{stochastic gradient descent} (SGD) algorithms. However, CLNs face critical issues of privacy violation, heavy communication overheads, and inscalability.

In view of the above issues, FL is proposed to enable distributed model training without a centralized training dataset \cite{yang2019federated}. As shown in Fig.\ref{fig:clnfln}b, a typical FLN is also composed of a server and multiple EDs. In FLNs, an ML model is trained by two iterative steps, namely the local model training at the EDs and the global model aggregation at the server. In the first step, the EDs update their local models with the global model downloaded from the server, execute SGD algorithms to train the local models with their own dataset, and upload the increments of the local models, i.e., model updates, to the server. In the second step, the server aggregates the received model updates by adding their average to the previous global model and obtains a new global model. The two steps constitute a training round. After multiple training rounds, the training will be completed once the global model converges. As such, in FLNs, the server and the EDs only exchange the ML model parameters, instead of raw data, which avoids the privacy issues and reduces the communication overheads significantly.

\subsection{Characteristics and Vulnerabilities of FLNs}\label{sec:FL_vul}

FLNs can be deployed flexibly in various environments, among which the most complicated one is the mobile implementation. For example, FLN has been adopted to orchestrate numerous mobile devices from all around the world to jointly train a language model for Gboard (https://ai.googleblog.com/2017/04/federated-learning-collaborative.html), while these mobile devices are owned by different users and connect to the server using different types of links, e.g., Wi-Fi and LTE. Therefore, the EDs in an FLN can be heterogenous in terms of ownership, computing capabilities, and connections \cite{mcmahan2016communication}. In addition, since FL relies on the joint effort of all EDs to train an ML model, the ML model will be tampered even if only few EDs work abnormally. Thus, FLNs have a broad attack surface. These characteristics, i.e., heterogeneity and broad attack surface, make FLNs vulnerable mainly from two aspects:

\begin{itemize}
  \item Malicious EDs: As smart devices are nowadays getting more sophisticated, flaws are inescapable and make the devices easily compromised by malwares. Meanwhile, a majority of existing FLN designs do not include an authentication mechanism, and thus they cannot prevent attackers from setting up malicious EDs to join FLNs.

  \item Insecure Connections: The EDs in an FLN may connect to the server via various connections. It is difficult to ensure the security of all the connections. For example, wireless connections are threatened by the openness of wireless channels. Although advanced encryption and verification methods can secure the connections, they induce additional overheads and thus are not always preferred, especially for some resource-limited IoT devices. Consequently, there may exist insecure connections, over which the downloaded global model or the uploaded model update can be hijacked and manipulated.

\end{itemize}

By exploiting the vulnerabilities of FLNs, attackers can inject poisoned model updates, which will tamper the global model aggregation and decrease the performance, i.e., accuracy, of the ML model. Such attacks are called poisoning attacks. Next, we describe the poisoning attacks on FLNs and existing countermeasures.

\section{Security Issues} \label{sec:app2}

\subsection{Poisoning Attacks}

Poisoning attacks aim to degrade an accuracy of the ML model by tampering the global model aggregation of FL with poisoned model updates.
According to the sources of poisoned model updates, poisoning attacks can be categorized into data poisoning and model poisoning \cite{lim2019federated}.

\subsubsection{Data Poisoning}

Data poisoning is carried out by means of modifying the training data in the compromised EDs. In particular, attackers flip the labels of training data, such that the compromised EDs train the local models using the poisoned data and generate incorrect model updates.
Depending on the attack intention, the labels can be flipped randomly or specifically. On the one hand, unintentional attacks target at decreasing the prediction accuracy on all classes, and thus attackers can flip the labels randomly \cite{munoz2019byzantine,xie2019slsgd,fu2019attack}. On the other hand, intentional attacks intend to make the ML model achieve a low accuracy on only certain classes, for which reason attackers only flip the labels of the training data in the concerned classes \cite{fu2019attack,fung2018mitigating}.

\subsubsection{Model Poisoning}

Instead of modifying training data, model poisoning produces poisoned model updates directly according to some pre-defined rules. For example, \cite{dong2019secure} considers that poisoned model updates can be sampled from a Gaussian distribution. Moreover, attackers can manipulate benign model updates into poisoned ones. In \cite{li2019rsa,dong2019secure,li2020learning}, malicious EDs are designed to flip the sign of benign model updates, in order to guide the aggregated global model towards the direction of decreasing accuracy. Similarly, \cite{fang2019local} designs poisoned model updates as the negative increment of the global model, leading to the reverse update of the global model. In fact, all the above attacks are unintentional as defined before. Alternatively, intentional model poisoning methods are considered in \cite{fu2019attack,li2020learning} and \cite{sun2019can}, where attackers use a pre-designed compromised model to craft poisoned model updates, aiming to replace the training ML model with the compromised model.

In addition, a scale factor is introduced in \cite{fu2019attack,dong2019secure}, and \cite{sun2019can} to magnify poisoned model updates, with the objective to further amplify attack effects.

\subsection{Countermeasures}

In the literature, there mainly exist three types of countermeasures to mitigate the poisoning attacks on FLNs, namely the robust aggregation methods, the anomaly detection-based methods, and the hyper methods.

\subsubsection{Robust Aggregation}

In FLNs, the server aggregates the received model updates by taking the average, allowing poisoned model updates to bias the global model directly. Hence, it is highly demanded to develop the aggregation methods that are robust to poisoned model updates. COMED, GEOMED, and COTMED are the commonly-used robust aggregation methods, which are proposed to replace the average operation with component-wise median, geometric median, and component-wise trimmed median, respectively \cite{xie2019slsgd,dong2019secure}. Another method called KRUM is proposed in \cite{dong2019secure}, which updates the global model by only using the most representative model update, i.e., the one with the shortest Euclidean distances from others. In \cite{sun2019can}, each model update is preprocessed to be within a bounded norm to prevent the global model from being overwhelmed by only few poisoned model updates. For the same purpose, \cite{li2019rsa} proposes a \emph{robust stochastic aggregation} (RSA) method, in which the server binarizes the received model updates before updating the global model.

\subsubsection{Anomaly Detection}

Benign and poisoned model updates have different objectives, making them implicitly different in mathematics. The anomaly detection-based methods aim to classify model updates by identifying the differences among them. For example, \cite{munoz2019byzantine} and \cite{li2020learning} propose to analyze model updates respectively by calculating their cosine similarities and by mapping them into a low-dimensional latent space. Then, the outliers, i.e., poisoned model updates, can be found and removed based on the obtained cosine similarities or the mapped low-dimensional representations.

\subsubsection{Hyper}

The hyper methods are proposed by combining the above two types of methods to put their merits into full use. In particular, the server needs to first evaluate the received model updates and then aggregate them in a robust way. For example, in \cite{fang2019local}, COMED, COTMED, and KRUM are enhanced with a preliminary evaluation procedure, where a model update will be discarded if it achieves an unacceptable loss or accuracy on an auxiliary dataset. Instead of simply discarding model updates, the methods proposed in \cite{fu2019attack} and \cite{fung2018mitigating} choose to reweight model updates based on the evaluation results. In \cite{fu2019attack}, the server analyzes model updates by using a repeated median estimator, and builds an accumulated confidence record for each ED. According to the confidence records, model updates are reweighted for aggregation. In contrast, the method in \cite{fung2018mitigating} is designed to reweight model updates based on their cosine similarities.

\subsection{Discussions and Open Issues}

\begin{table*}[t]
\centering
\caption{Summary of poisoning attacks and countermeasures.}
\includegraphics[width=16cm]{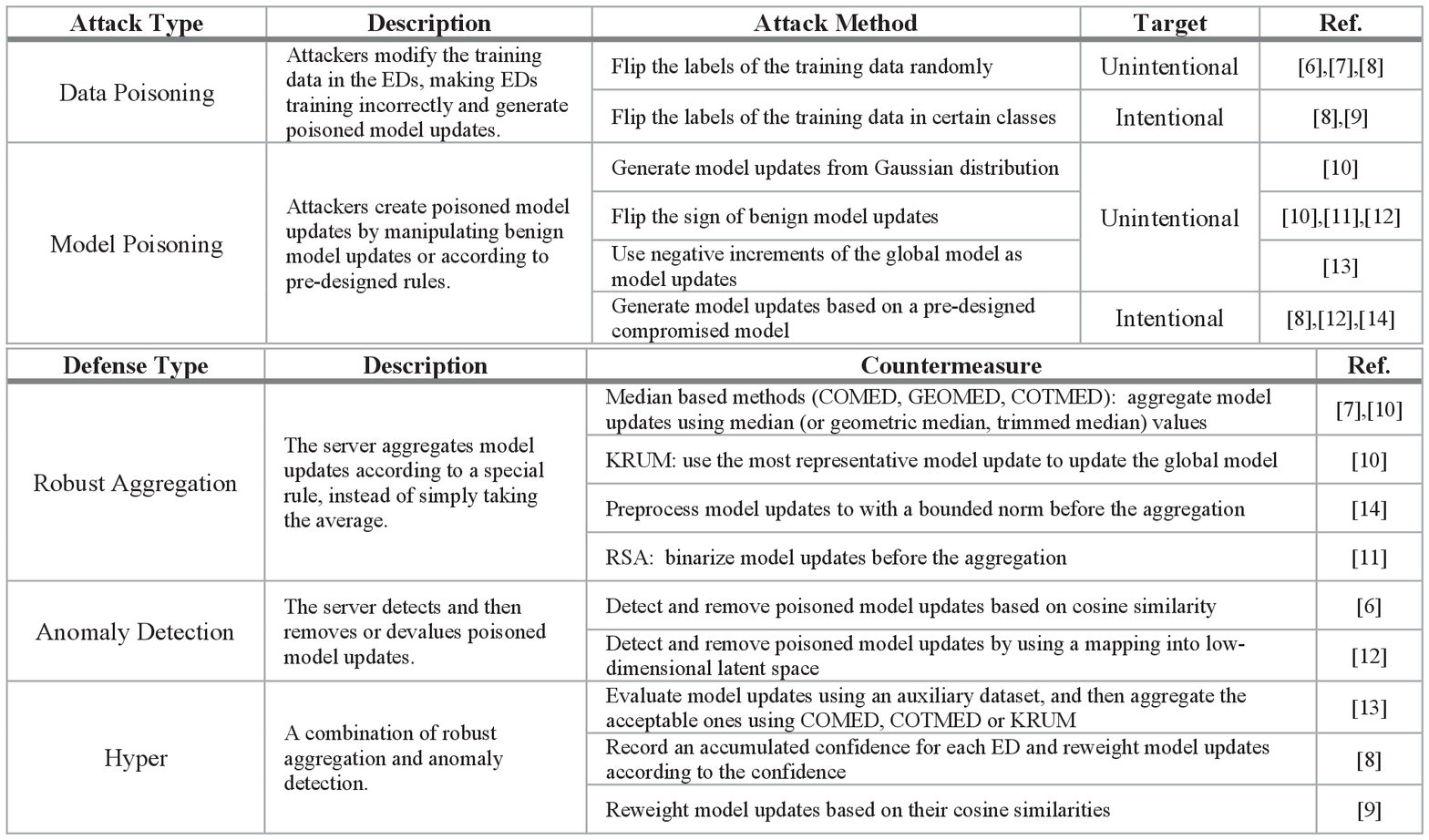}\label{tab:aio}
\end{table*}

We summarize the poisoning attacks and countermeasures in Table \ref{tab:aio}. From the table, the existing countermeasures are designed to prevent the global model from aggregating poisoned model updates, by way of discarding or devaluing part of the received model updates.
For example, in KRUM, only the most representative model update can be used to update the global model, while the others are all dropped.
In this sense, the existing countermeasures under-utilize model updates.

The under-utilization issue can be negligible if the server can obtain model updates for free. However, in practice, it is not guaranteed that the EDs naturally volunteer to contribute because they may not be interested in the model learnt by the server. Instead, as the owners of many valuable data, the EDs need to consume both computation and communication resources if they are asked to join the FL process. Hence, it is more practical to consider the existence of an incentive mechanism in FLNs, i.e., the EDs charge the server with training fees for contributing model updates \cite{lim2019federated}. To the best of our knowledge, there is a lack of effective security schemes that can deal with poisoning attacks while maintaining low training costs, which motivates us to develop a smart security enhancement framework.

\section{A Smart Security Enhancement Framework for Federated Learning Networks}

\subsection{FLN with Poisoning Attacks} \label{sec:model}

\begin{figure*} [t]
\centering
  \includegraphics[width=16cm]{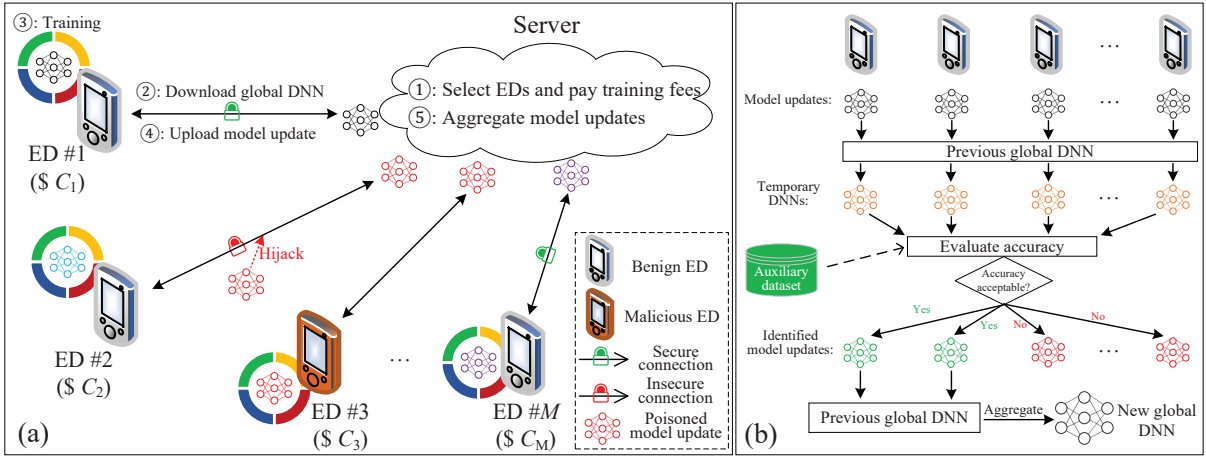}
  \caption{Illustrations of (a) system model of the considered FLN with $M$ EDs, (b) VBA procedure.}\label{fig:sysmodel}
\end{figure*}

In this paper, we consider an FLN with a server and $M$ EDs, as shown in Fig. \ref{fig:sysmodel}a. Particularly, the server pays and orchestrates the EDs to train ML models. Without loss of generality, the ML model is assumed to a DNN. Training the DNN requires multiple training rounds to complete. As an ED needs to invest resources once requested to participate in a training round, the EDs on request will charge the server with training fees before training local DNNs and uploading model updates. The details of each training round are shown in Fig. \ref{fig:sysmodel}a. In the FLN, attackers can exploit the malicious EDs and the insecure connections to launch poisoning attacks.
We refer the malicious EDs and the benign EDs with insecure connections to as vulnerable EDs, which can inject poisoned model updates if they are attacked successfully. In contrast, the benign EDs with secure connections are referred to as secure benign EDs, which are free from attacks and always contribute benign model updates.

To handle poisoned model updates, we propose a VBA procedure. As Fig. \ref{fig:sysmodel}b shows, the server first obtains temporary DNNs by adding each model update one at a time to the previous global DNN. Then, the temporary DNNs are tested on an auxiliary dataset.
If a temporary DNN can achieve acceptable accuracy, i.e., does not decrease the accuracy of the previous global DNN by a threshold, the corresponding model update is identified to be benign. The threshold exists to deal with the fact that a benign model update may still decrease the accuracy mildly due to natural fluctuations, and it can be designed empirically to be a small value slightly larger than zero.
Finally, the server updates the global DNN with the identified benign model updates.

In principle, the VBA procedure leverages the auxiliary dataset to determine whether a model update is poisoned or not. Thus, a proper auxiliary dataset should be representative of the pattern that the DNN learns, such that the accuracy degradations caused by poisoned model updates can be detected. We notice that the server typically has a set of test data, which are generally provided by the task publisher to evaluate the learnt DNN. Hence, we can use the test dataset as the auxiliary dataset. Note that although the VBA procedure is designed for poisoning attacks, it can be enhanced with extra methods to handle extensive cases. For example, lazy EDs may simply upload historical global model increments to cheat training fees. In this case, the VBA procedure can be additionally enhanced by checking the similarities between the received model updates and the historical global model increments. Since the server knows all the historical global model increments, the lazy EDs can be detected effectively as long as they copy any of them, even at the end of training. In fact, the lazy EDs tend to copy the recent global model increments to avoid destroying system performance, and thus the server only needs to store and use a few global model increments in the similarity check.

With the VBA procedure, poisoned model updates can be dropped, but the corresponding EDs have already been paid. Thus, it is highly desired to actively select appropriate EDs for each training round, with the objective to obtain the most benign model updates at the least training costs (i.e., the total training fees paid to the EDs). However, the server does not have enough information to select those appropriate EDs at the beginning of a training round. The reasons are two-fold. First, the behavior of vulnerable EDs is determined by attackers. Due to the malicious intention, attackers are impossible to inform the up-coming attacks in advance. Second, a model update cannot be recognized to be benign or not until it has been uploaded, while only the selected EDs can upload model updates in each training round. In other words, the server can only have a partial and historical observation of the EDs. Fortunately, the emergence of DRL makes it possible to make decisions without sufficient information. Next, we develop a smart ED selection strategy based on DRL, which empowers the server to learn the behavior of the EDs and to select EDs properly for each training round.

\subsection{A DRL-based ED Selection Strategy}

DRL is an important ML technique developed for decision-making in a dynamic environment \cite{mnih2015human,luong2019applications}.
Specifically, the decision maker, called agent, can learn environmental patterns and an optimal decision-making policy without requiring prior knowledge about the environment.
Hence, if we respectively model the server and the EDs as the agent and the environment, DRL can be used to enable the server to learn the behaving patterns of the EDs and to select the proper EDs, despite lacking sufficient prior knowledge.

To apply DRL, the interactions between the server and the EDs should be formulated as a \emph{Markov decision process} (MDP), which consists of three key elements, namely state, action, and reward.
At the beginning of a training round, a state is obtained by the server as the basis of decision-making, and thus it includes the status of all EDs, i.e., whether their previously uploaded model updates are benign or not.
Then, the server takes an action, i.e., selects a set of EDs.
After collecting all the model updates, the server will receive a reward to indicate how good the taken action is.
To encourage the server to obtain more benign model updates at lower training costs, we design the reward to be the number of received benign model updates minus the training costs, which is also the utility of the training round.

Solving the formulated MDP is to find the optimal ED selection policy that maximizes the long-term cumulative discounted reward.
DRL can solve the MDP effectively, of which the key idea is to establish a \emph{deep Q-network} (DQN) to approximate the Q-values for each state-action pair, i.e., the cumulative discounted rewards achieved by taking each action in each state \cite{luong2019applications}.
The DQN is a DNN containing an input layer, an output layer, and some hidden layers.
Given a specific state as the input, the DQN can predict the Q-value for each action.
To achieve accurate predictions, the DQN needs to be trained by a trial-and-error procedure, in which the agent generates experience by continuously interacting with the environment and feeding the recoded experience into the DQN.
The DQN will eventually converge after analyzing massive historical experience.
With the converged DQN, the server can always select the most appropriate EDs by choosing the action with the largest Q-value, as long as a state is given.
Note that we do not provide all technical details due to space limit. The interested readers can refer to \cite{luong2019applications} and \cite{mnih2015human} for more details about DRL.

Nevertheless, it is not the end of the story. Recall that only the EDs selected in a training round can upload model updates, while the status of an ED needs to be determined by running the VBA procedure with its model update received at the server. Therefore, the server can only have a partial observation of the state, which transforms the MDP into a \emph{partially observable Markov decision process} (POMDP) \cite{luong2019applications}. The information included in a single observation is incomplete, but fortunately we notice that the historical status of the unselected EDs can help to supplement the missing information. In other words, a state can be predicted from the present observation and some historical observations. Following this idea, we first define a pseudo state as a sequence of present and historical observations, and the corresponding taken actions. Next, we design a \emph{deep recurrent Q-network} (DRQN) \cite{hausknecht2015deep} by inserting a \emph{long-short-term-memory} (LSTM) layer into the vanilla DQN. LSTM is kind of DNN structure designed for analyzing sequential data, e.g., a period of sound waves. Since the pseudo state is exactly a time sequence of observations and actions, LSTM can be used to analyze the pseudo state and to predict the real state behind it. After the analysis of LSTM, a predicted state can be extracted from the pseudo state and be handled by the DQN.

\subsection{Numerical Results}

\begin{figure*} [t]
\centering
  \includegraphics[width=15cm]{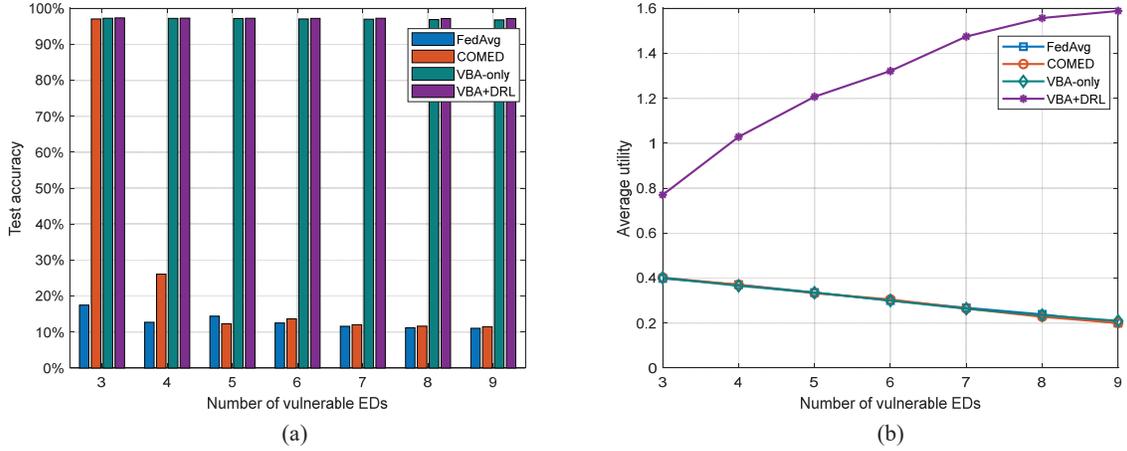}
  \caption{The test accuracy and average utility achieved by the FedAvg and COMED algorithms, and the proposed framework.}\label{fig:results}
\end{figure*}

\begin{figure*} [t]
\centering
  \includegraphics[width=16cm]{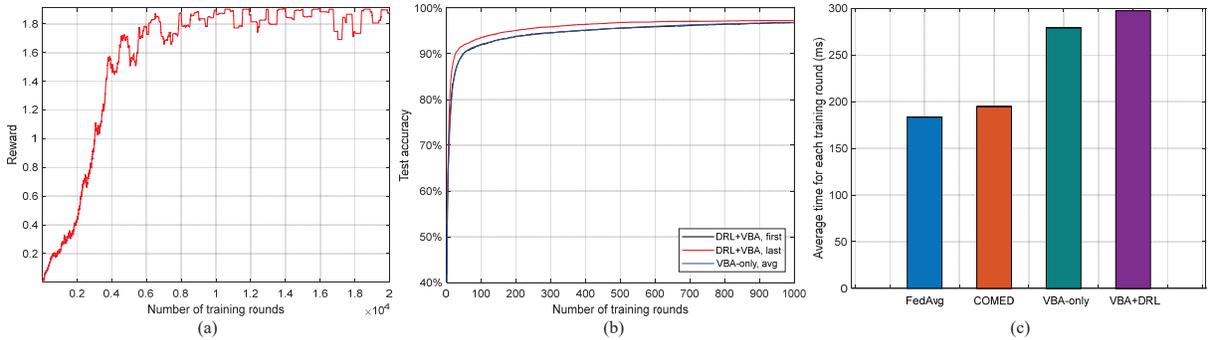}
  \caption{The efficiency of the proposed framework.}\label{fig:results2}
\end{figure*}

In this part, we conduct simulations to evaluate the proposed smart security enhancement framework in an FLN with ten EDs, including both secure benign EDs and vulnerable EDs. Since vulnerable EDs may upload poisoned model updates, they tend to set a lower price than that of secure benign EDs to attract more requests. In particular, the secure benign EDs and the vulnerable EDs are considered to respectively charge the price of $0.9$ and $0.3$ for each request. If a vulnerable ED is under attack, the labels of training data will be flipped randomly.
Moreover, a scale factor is imposed to magnify the poisoned model updates by $20$ times. Considering that attackers can launch attacks dynamically, the vulnerable EDs behave differently in each training round, i.e., each vulnerable ED can successfully upload a benign model update according to a hidden pattern.

We consider that $20$ learning tasks are published to the FLN, which we simulate by repeatedly training a DNN over an MNIST dataset with $60000$ training samples and $10000$ test samples. The training samples are distributed to each ED equally. The DNN to be learnt contains $2$ \emph{fully-connected} (FC) hidden layers with $100$ neurons in each layer. Each learning task consists of $1000$ training rounds, and five EDs will be selected in each round to train locally with the minibatch size of $100$ for $1$ epoch, using Adam as the SGD optimizer. As for the DRQN, it contains an LSTM hidden layer with $32$ LSTM units and an FC hidden layer with $200$ neurons. Each pseudo state contains three successive observations. The threshold in VBA is set to be $0.005$. Additionally, we take the FedAvg \cite{mcmahan2016communication} and the COMED algorithms as the benchmark algorithms. For the proposed framework, we consider the full implementation, i.e., ``VBA+DRL'', and the VBA-only implementation, where the VBA procedure works with random ED selection.

We evaluate and depict in Fig. \ref{fig:results} the performance of the four schemes by increasing the number of the vulnerable EDs from $3$ to $9$. Fig. \ref{fig:results}a shows the test accuracy of the DNNs after training, and each data point is the average result of all learning tasks. As seen, the test accuracy of the DNNs trained by the benchmark algorithms is degraded significantly due to attacks. In particular, the FedAvg algorithm is the worst because it does not have any robustness design. For the COMED algorithm, it can train DNNs to achieve an acceptable accuracy of around $96\%$ at the presence of three vulnerable EDs but fails to prevent severe degradation with more vulnerable EDs.
In contrast, with either implementation, the DNNs trained by the proposed framework can always reach an accuracy of over $97\%$, regardless of the number of the vulnerable EDs in the network. This demonstrates that the VBA procedure can protect the FLN from poisoning attacks effectively. Fig. \ref{fig:results}b shows the average utility of each training round. From the figure, the VBA-only implementation and the benchmark algorithms achieve lower utility with more vulnerable EDs. This is because they all select EDs randomly while the growth of vulnerable EDs increases the chance of obtaining poisoned model updates. Meanwhile, the full implementation of our proposed framework can achieve higher utility. The reasons are two-fold. First, more vulnerable EDs bring in more potential benign model updates charging low training fees. Second, enabled by DRL, the server can select EDs smartly after considering both attacks and training costs. Hence, compared with the random ED selection, the DRL-based one can obtain more low-cost and benign model updates.

Next, we look at the efficiency of the proposed framework by taking the case with $9$ vulnerable EDs for example. Fig. \ref{fig:results2}a shows the evolution of the rewards achieved by DRL for $20000$ training rounds in the full implementation, which cover all the $20$ learning tasks. As seen, DRL takes around $8000$ training rounds (i.e., $8$ learning tasks) to converge. Note that although DRL needs some time to converge, it can maintain high performance afterwards. Thus, the convergence time is tolerable from a long-term perspective of the FLN.
We further depict in Fig. \ref{fig:results2}b the impacts of DRL on the convergence speed of federated learning. In particular, the red curve shows the evolution of the test accuracy of the DNN trained in the last learning task, where DRL has converged, while the black curve shows that in the first learning task, where DRL just starts to learn and has not converged. For comparison, Fig. \ref{fig:results2}b also shows the evolution of the VBA-only implementation, and each data point is the average result of all learning tasks. As seen, the accuracy increases sharply and can converge at around $600$ training rounds for the last task. As for the first task, the convergence rate is slower and the curve is the same as that of the VBA-only implementation. Hence, as DRL is converging, FL can achieve better convergence performance.
Meanwhile, even before the convergence, the performance of the DRL-based ED selection is not worse than that of the random ED selection.
Finally, in Fig. \ref{fig:results2}c, we compare the four schemes regarding the average time for each training round. Compared with the benchmark algorithms, the proposed framework has higher latency due to the more complicated aggregation procedure. However, the VBA-only and the full implementations can still finish a training round respectively within $279$ms and $297$ms. In practice, the FLN can choose either implementation depending on the practical latency or computation requirements.

\section{Conclusions}\label{sec:conclusion}
As FLNs are susceptible to poisoning attacks, where attackers attempt to disrupt the training process of FL by injecting poisoned model updates, we have investigated the security issues of FLNs in this paper. In particular, we have first reviewed the vulnerabilities of FLNs, and then provided an overview of poisoning attacks and mainstream countermeasures. However, it has been found that the existing countermeasures fail to consider the costs of requesting EDs to contribute model updates and under-utilize model updates, leading to inefficient training. Therefore, we have proposed a smart security enhancement framework to address this issue. In the proposed framework, we have designed a VBA procedure for identifying and removing poisoned model updates and a DRL-based ED selection strategy for intelligently selecting the EDs that can provide low-cost and benign model updates. Numerical results have demonstrated that the proposed framework can protect FLNs effectively and efficiently.

\bibliographystyle{IEEEtran}

\end{document}